\documentclass[aps,prb,superscriptaddress,twocolumn]{revtex4-1}

\usepackage{todonotes}

\usepackage{times}
\usepackage{amsmath}
\usepackage{amsfonts}
\usepackage{amssymb}
\usepackage{epsfig}
\usepackage{siunitx}
\usepackage{gensymb}

\usepackage{float}
\usepackage{ifthen}
\usepackage{xspace}
\usepackage{relsize}

\usepackage{algorithm}
\usepackage{algorithmic}

\begin{document}

\title{Observation of edge magnetoplasmon squeezing in a quantum Hall conductor}
\author{ H. Bartolomei$^{1}$, R. Bisognin$^{1}$, H. Kamata$^{1}$,  J.-M. Berroir$^{1 }$,  E. Bocquillon$^{1,2 }$, G. M\'{e}nard$^{1 }$, B.
Pla\c{c}ais$^{1 }$,   A. Cavanna$^{3}$, U.
Gennser$^{3}$, Y. Jin$^{3}$,   P. Degiovanni$^{4}$, C. Mora$^{5}$, and G. F\`{e}ve$^{1 \ast}$ \\
\normalsize{$^{1}$ Laboratoire de Physique de l\textquoteright Ecole normale sup\'erieure, ENS, Universit\'e
PSL, CNRS, Sorbonne Universit\'e, Universit\'e Paris Cit\'e, F-75005 Paris, France}\\
\normalsize{$^{2}$ II. Physikalisches Institut, Universit\"{a}t zu K\"{o}ln, Z\"{u}lpicher Str. 77, 50937 K\"{o}ln}\\
\normalsize{$^{3}$ Centre de Nanosciences et de Nanotechnologies (C2N), CNRS, Universit\'{e} Paris-Saclay, 91120 Palaiseau, France. }\\
\normalsize{$^{4}$ Univ  Lyon,  Ens  de  Lyon,  Universit\'e  Claude  Bernard  Lyon  1, CNRS,  Laboratoire  de  Physique,  F-69342  Lyon,  France}\\
\normalsize{$^{5}$ Universit\'e  de  Paris,  Laboratoire  Mat\'eriaux  et  Ph\'enom\`enes  Quantiques,  CNRS,  F-75013  Paris,  France.}\\
\normalsize{$^\ast$ To whom correspondence should be addressed; E-mail:  gwendal.feve@ens.fr.}\\}

\begin{abstract}
Squeezing of the quadratures of the electromagnetic field has been extensively studied in optics and microwaves. However, previous works focused on the generation of squeezed states in a low impedance ($Z_0 \approx 50 \Omega$) environment. We report here on the demonstration of the squeezing of bosonic edge magnetoplasmon modes in a quantum Hall conductor whose characteristic impedance is set by the quantum of resistance ($R_K \approx 25 k \Omega$), offering the possibility of an enhanced coupling to low-dimensional quantum conductors. By applying a combination of dc and ac drives to a quantum point contact, we demonstrate squeezing and observe a noise reduction 18\% below the vacuum fluctuations. This level of squeezing can be improved by using more complex conductors, such as ac driven quantum dots or mesoscopic capacitors.
\end{abstract}
\maketitle

In quantum Hall conductors, charge excitations propagate ballistically along one dimensional chiral channels. This ballistic propagation has been exploited in electron quantum optics experiments \cite{Bocquillon2014,Marguerite2017} focusing on the generation and manipulation of elementary electron and hole excitations of the Fermi sea. These are particle-like fermionic excitations, but the dynamics of charge propagation along one-dimensional edge channels can be equivalently described in terms of collective bosonic excitations called edge magnetoplasmons (EMP), which consist of coherent superpositions of electron-hole pairs on top of the Fermi sea.

EMP have been largely investigated in the past by studying the propagation of the time dependent electrical current in the time \cite{Ashoori1992,Zhitenev1993,Ernst1996,Sukhodub2004,Kamata2010,Kumada2011} or frequency domain\cite{Talyanskii1992,Gabelli2007a,Hashisaka2012,Delgard2021}. Experiments have highlighted the dependence of EMP propagation speed on the magnetic field and on the screening by nearby electrostatic gates. All these studies are based on a classical description of charge propagation along the edge channels which can be modeled as transmission lines\cite{Hashisaka2012}. However, chiral edge channels have three important differences with respect to standard 50 Ohms coaxial cables. Firstly, the chirality results in the separation between forward and backward propagating waves. Secondly, the speed of the EMP\cite{Kumada2011} is of the order of $10^5$m.s$^{-1}$, three orders of magnitude smaller than the speed of light, resulting in wavelengths in the $\mu$m range at GHz frequencies compared to the cm range in standard coaxial cables. EMPs would thus allow for more compact circuits. Finally, their characteristic impedance is of the order of the resistance quantum, $R_K \approx 25$ k$\Omega$, much larger than the 50 Ohms standard, offering the possibility of a strong coupling to low dimensional quantum conductors of high impedance\cite{Parmentier2011b}.

These specificities motivated recent theoretical and experimental \cite{Viola2014} \cite{Bosco2019} studies of EMP transmission lines for efficient coupling to on-chip high impedance quantum devices, such as charge or spin qubits, for the study of Coulomb interaction effects in one-dimensional edge channels \cite{Gourmelon2020}, or for the realization of on-chip microwave circulators \cite{Mahoney2017}. So far, these studies have focused on the classical regime, where EMP states can be described as coherent states. However, as for other bosonic modes, quantum EMP states can also be generated. In the last years there has been a strong interest for the generation of quantum radiation by quantum conductors \cite{Beenakker2004,Lebedev2010,Grimsmo2016} and in particular of squeezed states \cite{Mendes2015,Ferraro2018,Rebora2021}. So far, it has been limited to the study of  low impedance (50 Ohms) transmission lines coupled to superconducting circuits\cite{Armour2013,Mendes2019,Rolland2019} or tunnel junctions\cite{Gasse2013}. We report here on the generation of squeezed EMP states at the output of a quantum
point contact used as an electronic beam-splitter in a GaAs quantum Hall conductor, as discussed in Ref. [\onlinecite{Rebora2021}]. Using two-particle interference processes\cite{Marguerite2017} occurring between electron and hole excitations colliding on the splitter, we generate a squeezed EMP vacuum state at frequency $f=\frac{\Omega}{2\pi}=7.75$ GHz at the splitter output with a noise minimum 18\% below the vacuum fluctuations. The non-linear EMP scattering at the splitter breaks a $2f$ pump signal into coherent photon pairs, thereby achieving squeezing \cite{Yuen76}.

Squeezed EMP states could be used for quantum enhanced measurements in EMP interferometers\cite{Hiyama2015}, or to extend the study of low dimensional quantum conductors in the regime where they are driven by quantum voltage sources\cite{Souquet2014}, exploiting the strong coupling of high impedance transmission lines to high impedance low-dimensional quantum circuits.

In the bosonic description of charge propagation, the charge density $\rho(x,t)$ carried by a single edge channel can be expressed as a function of a chiral bosonic field $\Phi(x,t)$ with $\rho(x,t)=\frac{-e}{\sqrt{\pi}}\partial_x \Phi(x,t)$. The relation between the electrical current and the field can then be deduced directly from charge conservation: $i(x,t)=\frac{e}{\sqrt{\pi}}\partial_t \Phi(x,t)$. At low frequency (typically a few GHz), dispersion effects can be neglected, such that $\Phi(x,t)$ can be decomposed in terms of elementary plasmon excitations at pulsation $\omega$ propagating with constant speed velocity $v$:
\begin{eqnarray}
\Phi(x,t) & = & \frac{-i}{\sqrt{4\pi}} \sum_{\omega} \sqrt{\frac{2\pi}{\omega T_{\text{meas}}}} [b_{\omega} e^{i \omega (x/v-t)} -h.c.] \\
\text{with} & & [b_{\omega}, b^{\dag}_{\omega'}]=\delta_{\omega, \omega'}.
\end{eqnarray}
$b^{\dag}_{\omega}$ is the operator which creates a single plasmon of energy $\hbar \omega$ and obeys the usual bosonic commutation relations. The long measurement time $T_{\text{meas}}$ sets the discretization of the plasmon modes by steps of  $2\pi/T_{\text{meas}}$. In order to address the squeezing of EMP modes, it is useful to introduce the quadratures of the bosonic field at a given pulsation $\Omega$ defined for a phase $0\leq \varphi \leq \pi$:
\begin{eqnarray}
X_{\Omega, \varphi} &= & \frac{b_{\Omega} e^{i \varphi} + b^{\dag}_{\Omega} e^{-i \varphi}}{\sqrt{2}}
\end{eqnarray}
Their fluctuations $\langle \Delta X_{\Omega, \varphi}^{2}\rangle$ can be decomposed into an isotropic $\langle \Delta X_{\Omega, \text{iso}}^{2}\rangle$ and an anisotropic, $\varphi$ dependent, part $\langle \Delta X_{\Omega, \text{an}}^{2}\rangle$ given by:
\begin{eqnarray}
\langle \Delta X_{\Omega, \text{iso}}^{2}\rangle & = & \langle b^{\dag}_\Omega b_\Omega \rangle -\langle b^{\dag}_\Omega \rangle \langle b_\Omega \rangle +\frac{1}{2}   \\
\langle \Delta X_{\Omega, \text{an}}^{2}\rangle & =& \Re\left((\langle b^{2}_\Omega \rangle -\langle b_\Omega \rangle ^{2}) e^{2 i \varphi} \right)
\end{eqnarray}
For classical coherent states, $\langle \Delta X_{\Omega, \varphi}^{2}\rangle$ is isotropic ($\langle \Delta X_{\Omega, \text{an}}^{2}\rangle=0$) and given by $\langle \Delta X_{\Omega, \varphi}^{2}\rangle =1/2$ which are called vacuum fluctuations. For squeezed states, the minimum value of the noise, obtained for a certain value $\varphi= \varphi_0$ of the angle, goes below the vacuum fluctuations, $\langle \Delta X_{\Omega, \varphi_0}^{2}\rangle <1/2$. As imposed by Heisenberg's uncertainty principle, the orthogonal quadrature then exhibits larger fluctuations: $\langle \Delta X_{\Omega, \varphi_0 + \pi/2}^{2}\rangle >1/2$.

The quadratures of the field and their fluctuations can be experimentally accessed from the measurements of the electrical current $i(t)$ and its fluctuations at high frequency. More precisely, defining $i_{\Omega,\varphi}(t)  =  \cos({\Omega t + \varphi}) \; i(t)$, one has:
\begin{eqnarray}
\langle \overline{ i_{\Omega,\varphi}(t)}^{T_{\text{meas}}} \rangle & = & -2e \sqrt{\frac{ \Omega}{\pi T_{\text{meas}}}} \langle X_{\Omega,\varphi} \rangle \\
S_{\Omega,\varphi} & = &2 \int d\tau \overline{\langle \delta  i_{\Omega,\varphi}(t+ \tau/2) \delta i_{\Omega,\varphi}(t -\tau/2) \rangle}^{T_{\text{meas}}} \\
 & =& \frac{e^2 \Omega}{2\pi} \langle \Delta X_{\Omega,\varphi}^{2} \rangle \label{X2}
\end{eqnarray}
where $\overline{ i_{\Omega,\varphi}(t)}^{T_{\text{meas}}}$ denotes the average of $i_{\Omega,\varphi}(t)$ over the measurement time $T_{\text{meas}}$. Classical states are thus defined by current fluctuations $S_{\Omega,\varphi}= \frac{e^2 \Omega}{4\pi}= \frac{e^2 f}{2}$ and squeezed states by $S_{\Omega,\varphi_0}< \frac{e^2 f}{2}$. Experimentally, one measures the noise in excess of the equilibrium fluctuations $\Delta S_{\Omega,\varphi}= S_{\Omega,\varphi} -\frac{e^2 f}{2}$ and squeezing occurs when $\Delta S_{\Omega,\varphi_0} <0$ \cite{Note1}.

\begin{figure}[h!]
\includegraphics[width=1
\columnwidth,keepaspectratio]{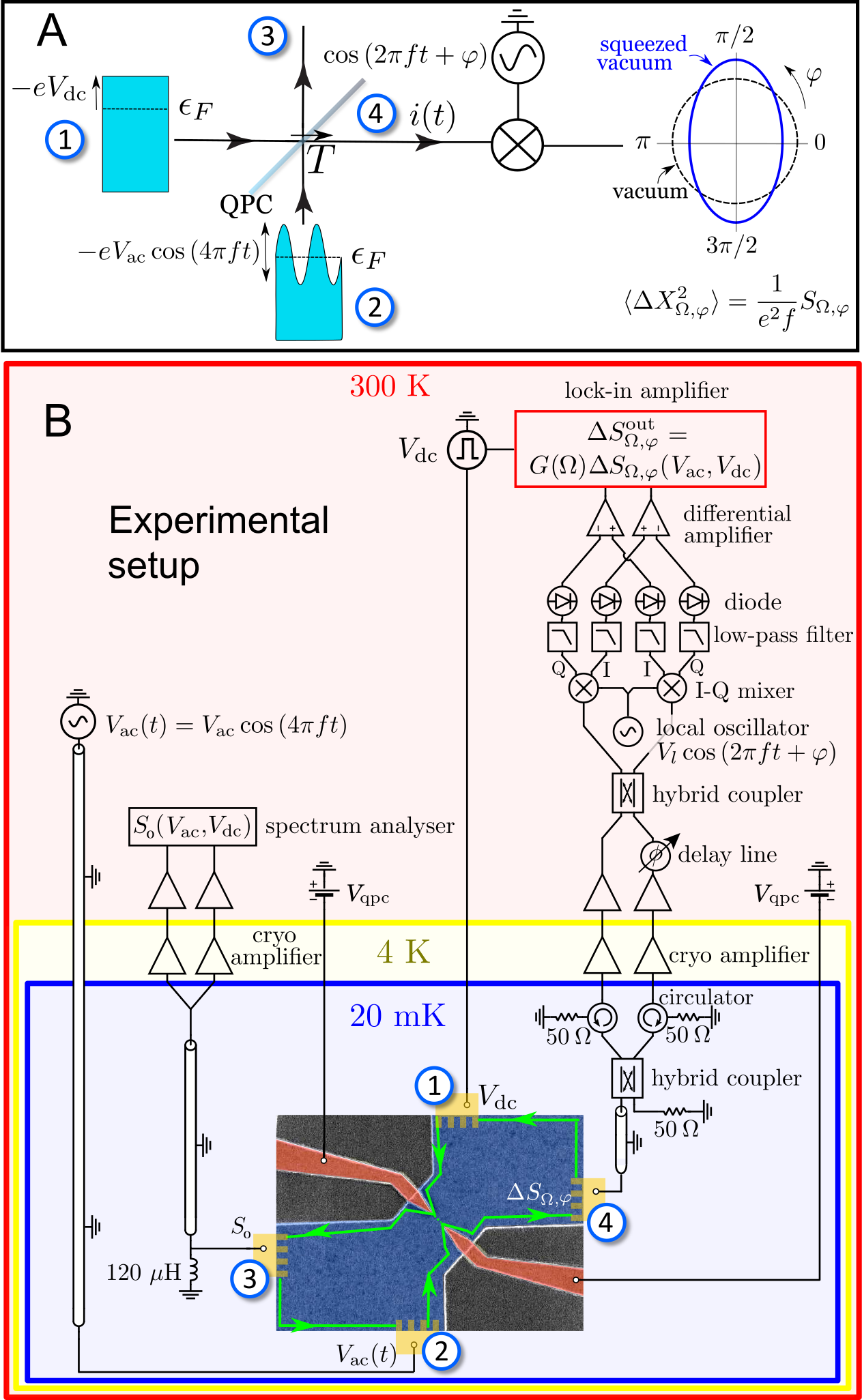}
\caption{\label{fig1}
{\bf A.} Principle of the experiment: a QPC is used as an electronic beam-splitter of transmission $T$ for the collision of electron and hole excitations generated at inputs 1 and 2. Input 1 is connected to a dc source, which shifts the chemical potential of the edge channel by $-eV_{\text{dc}}$. Input 2 is connected to an ac sinusoidal source of amplitude $V_{\text{ac}}$ and frequency $2f$. EMP squeezing is characterized at output 4 by measuring the correlations $S_{\Omega,\varphi}$ of the current $i_{\Omega,\varphi}$. For a squeezed vacuum, $S_{\Omega,\varphi}$ goes below the vacuum fluctuations for $\varphi=0$ and above them for $\varphi=\pi/2$. {\bf B.} Experimental setup: the low frequency noise $\Delta S_0$ is measured at output 3 and the high frequency noise $\Delta S_{\Omega,\varphi}$ at output 4. The EMP current $i(t)$ is weakly transmitted to a $Z_{0}=50$ $\Omega$ coaxial cable, amplified using two cryogenic amplifiers in a double balanced configuration\cite{Parmentier2011,Bisognin2019b} and multiplied to the local oscillator $V_{l}(t)=V_l \cos{(2\pi f t + \varphi)}$ using  high frequency mixers. The noise power is then integrated using a diode in a $800$ MHz bandwidth set by a low pass filter. The noise is measured via a lock-in detection by applying a square modulation at a frequency $f_m=234$ Hz to the dc voltage $V_{\text{dc}}$. The resulting output excess noise, $\Delta S_{\Omega,\varphi}^{\text{out}}$ is proportional to $\Delta S_{\Omega,\varphi}$ with a proportionality factor $G(\Omega)$ that needs to be calibrated. }
\end{figure}

The principle of the experiment is represented in Figure 1.A. A quantum point contact (QPC) is used as a beam-splitter for electronic excitations of transmission probability $T$ (and reflection probability $R=1-T$). By plugging two electronic sources at inputs 1 and 2 of the QPC, collisions between electron and hole excitations\cite{Liu1998,Bocquillon2013} emitted by each source occur at the beam-splitter. Previous implementations of high frequency noise measurements \cite{Zakka-Bajjani2007,Bisognin2019b} have focused on the single source configuration, where the quantum point contact is biased by a dc voltage $V_{\text{dc}}$. These measurements have shown that there is a voltage threshold for the generation of high frequency noise $V_{\text{dc}}=hf/q$ which can be used to measure the charge $q$ of the quasiparticles scattered by the splitter\cite{Bisognin2019b}. However, in this configuration, the noise is isotropic and no squeezing is obtained. By adding an ac sinusoidal source at frequency $2f$ at input 2, EMP squeezing at frequency $f$ can be obtained when both the dc voltage $V_{\text{dc}}$ and the ac voltage amplitude $V_{\text{ac}}$ are set close to the threshold $hf/q$. Squeezing is characterized by multiplying the electrical current $i(t)$ at the output of the splitter with the local oscillator $\cos({\Omega t + \varphi})$. As represented in Figure 1.A., the resulting low-frequency current correlations $S_{\Omega,\varphi}$ are expected to go beyond (respectively above) the vacuum fluctuations when the phase of the local oscillator is set to $\varphi=0$ (respectively $\varphi=\pi/2$). This approach has strong similarities with the one developed by Gasse et al. in Ref. [\onlinecite{Gasse2013}] where squeezing is generated in a $50$ Ohms coaxial line using a low impedance tunnel junction of resistance $R=70$ $\Omega$. As mentioned above, our work is different as it demonstrates squeezing in a high impedance $Z\approx R_K$ EMP transmission line allowing for a strong in situ coupling to mesoscopic circuits.

The experimental setup is represented in Figure 1.B. The conductor is a two-dimensional electron gas in a GaAs/AlGaAs heterostructure, with charge density $\SI{1.9e15}{\per\square\meter}$ and mobility
$\mu=\SI{2.4e6}{\per\square\centi\meter\per\volt\per\second}$ at 4 K. A magnetic field $B=2.6$ T is applied perpendicularly to the sample in order to reach the quantum Hall effect at filling factor $\nu=3$ where three edge channels propagate along the edges of the sample. A quantum point contact is used to partition selectively the outer edge channel on which we focus in this study. The transmission is set to $T \approx 0.5$ to generate the maximum partition noise.  A dc current is generated at input 1 of the quantum point contact and the ac voltage $V_{\text{ac}}(t)=V_{\text{ac}}\cos{(4\pi f t)}$ is applied at input 2 with $f=7.75$~GHz. The measurement frequency $f=7.75$~GHz is chosen such that $k_B T_{\text{el}} \ll hf$, where $T_{\text{el}}$ is the electronic temperature. The high frequency noise $\Delta S_{\Omega,\varphi}(V_{\text{dc}},V_{\text{ac}})$ is measured by weakly transmitting the EMPs propagating at output 4 to a 50 Ohm coaxial line, where the weak coupling is ensured by the strong impedance mismatch between the $50$ Ohms of the coaxial cable and the impedance $R_K/\nu$ of the quantum Hall conductor. The signal is then amplified by a set of two double balanced cryogenic amplifiers\cite{Parmentier2011,Bisognin2019b}. $S_{\Omega,\varphi}$ is measured by multiplying the output signal with a local oscillator $V_{l}(t)=V_{l}\cos{(\Omega t + \varphi)}$ using high frequency mixers. The local oscillator is locked in phase with the pump $V_{\text{ac}}(t)$ and the phase $\varphi$ of the measured quadrature can be continuously varied. $S_{\Omega,\varphi}$ is finally measured using a diode which integrates the power at the output of the mixer in a $800$ MHz bandwidth set by a low pass filter.
The typical resolution required on $S_{\Omega,\varphi}$ is smaller than $10^{-29}$ A$^2$ Hz$^{-1}$, which corresponds to the thermal noise generated by a variation of a few tens of microKelvins of a $50$ Ohm resistor. This needs to be compared to the base temperature of the fridge ($\approx 30$ mK) and to the noise temperature of the cryogenic amplifiers ($\approx 4$ K). In order to mitigate longtime variations of these two noise temperatures which could easily overcome the signal, we use a lock-in detection of the noise by modulating in time the applied dc voltage on input 1 (between $V_{\text{dc}}$ and 0) at a frequency $f_m=234$ Hz. The output excess high frequency noise $\Delta S^{\text{out}}_{\Omega,\varphi}(V_{\text{dc}},V_{\text{ac}})$  is proportional to $\Delta S_{\Omega,\varphi}(V_{\text{dc}},V_{\text{ac}})$ with a proportionality factor $G(\Omega)$ that takes into account both the weak coupling to the transmission line and the amplification chain.

\begin{figure}[h!]
\includegraphics[width=1
\columnwidth,keepaspectratio]{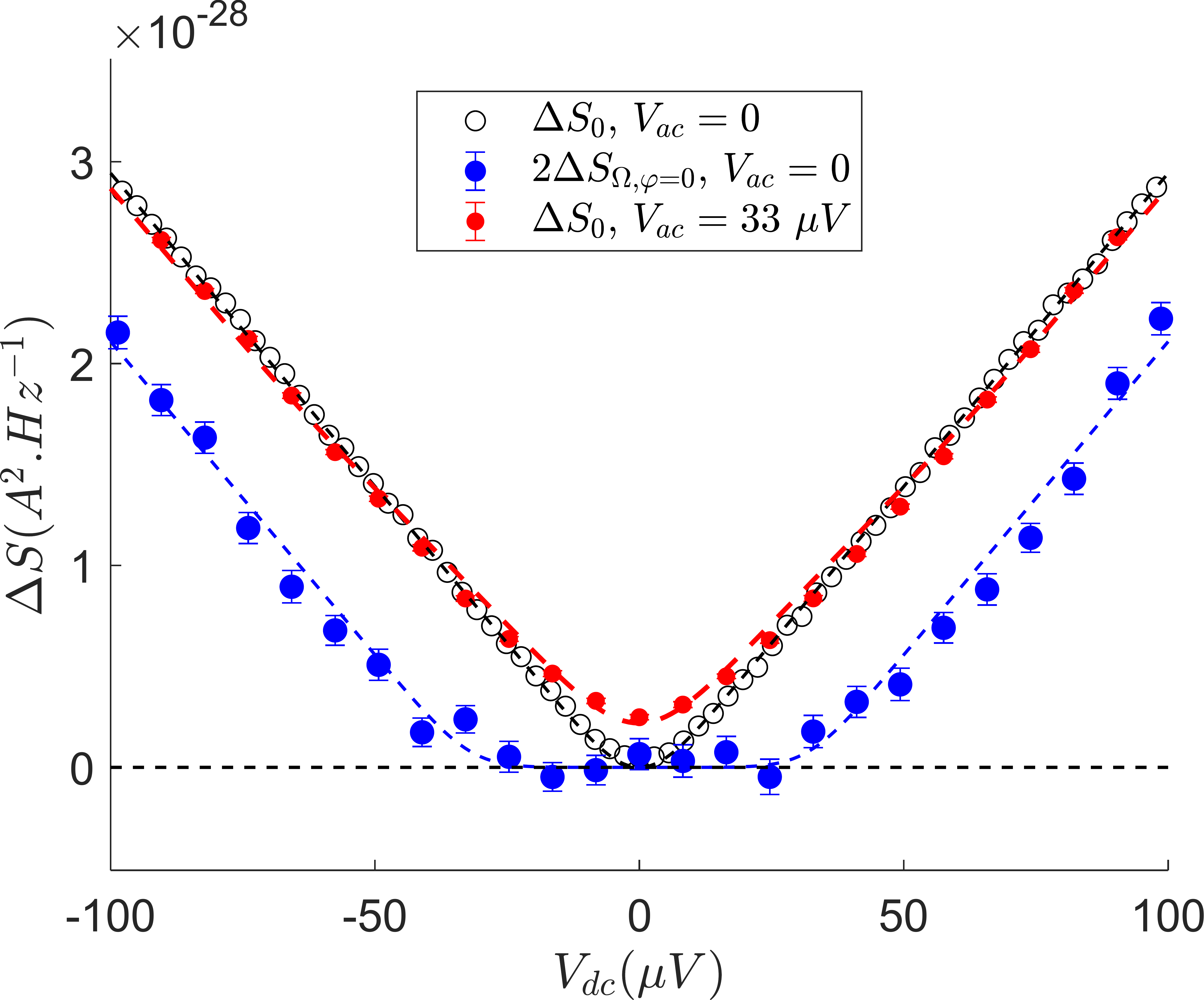}
\caption{\label{fig2}
Measurements of $\Delta S_0(V_{\text{dc}},V_{\text{ac}}=0)$ (black circles) and $\Delta S_{\Omega_\varphi}(V_{\text{dc}},V_{\text{ac}}=0)$ (blue points) when the pump is switched off, and measurements of $\Delta S_0(V_{\text{dc}},V_{\text{ac}})$ when the pump is switched on with $V_{\text{ac}}=33$ $\mu$V (red points). The black and red dashed lines represent the theoretical predictions for $\Delta S_0$ with $V_{\text{ac}}=0$ (black dashed line) and $V_{\text{ac}}=33$ $\mu$V (red dashed line) with $T_{\text{el}}=40$ mK. The blue dashed line is the theoretical prediction for $\Delta S_{\Omega,\varphi}$ with $V_{\text{ac}}=0$ and $T_{\text{el}}=30$ mK.}
\end{figure}
We calibrate $G(\Omega)$ by measuring simultaneously the easily calibrated \cite{Note2} excess zero frequency noise $\Delta S_{0}(V_{\text{dc}},V_{\text{ac}}=0)$ (at output 3 of the splitter) and the high frequency noise $\Delta S_{\Omega,\varphi}(V_{\text{dc}},V_{\text{ac}}=0)$ (at output 4) when the pump is off, $V_{\text{ac}}=0$.  In the absence of the pump, $\Delta S_{\Omega,\varphi}(V_{\text{dc}},0)$ is independent of the phase $\varphi$ and is directly related to the excess zero frequency noise shifted by the voltage\cite{Roussel2016,Safi2020} $\pm hf/e$:
\begin{eqnarray}
 \Delta S_{\Omega,\varphi}(V_{\text{dc}},0) & = & \frac{\Delta S_0(V_{\text{dc}}+hf/e) + \Delta S_0(V_{\text{dc}}-hf/e)}{4} ,\label{RFLF}
\end{eqnarray}
with $\Delta S_0(V_{\text{dc}} \pm hf/e)= S_0(V_{\text{dc}} \pm hf/e) -S_0(\pm hf/e)$. As can be seen from Eq. (\ref{RFLF}), for high voltages $V_{dc}> hf/e$, $ \Delta S_{\Omega,\varphi}(V_{\text{dc}},0)$ varies linearly with the applied voltage with a slope $\frac{e^3}{h} R T$ which can be used for the determination of $G(\Omega)$. The excess low frequency noise $\Delta S_0(V_{\text{dc}},0)$ and the high frequency noise $2 \Delta S_{\Omega,\varphi=0}(V_{\text{dc}},0)$ for $f=\frac{\Omega}{2\pi}=7.75$ GHz are plotted in Figure 2, where the constant $G(\Omega)$ has been adjusted so that $\Delta S_{\Omega,\varphi}(V_{\text{dc}},0)$ verifies Eq. (\ref{RFLF}). The black dotted line represents the standard low-frequency shot noise formula $\Delta S_0(V_{\text{dc}},0)=2 \frac{e^3}{h} R T V_{\text{dc}} [\coth{\frac{eV_{\text{dc}}}{2k_B T_{\text{el}}}}-\frac{2k_B T_{\text{el}}}{eV_{\text{dc}}}]$ with $T_{\text{el}}=40$ mK and the transmission of the beam-splitter is set to $T=0.42$. The blue dashed line represents the prediction from Eq.(\ref{RFLF}) which agrees well with the data. As expected, the low frequency noise varies linearly for $eV_{\text{dc}}>k_BT_{el}$ which contrasts with the high frequency noise which is suppressed when $V_{\text{dc}}$ is smaller than the voltage threshold $hf/e =32$ $\mu$V for $f=7.75$ GHz.

\begin{figure}[h!]
\includegraphics[width=1
\columnwidth,keepaspectratio]{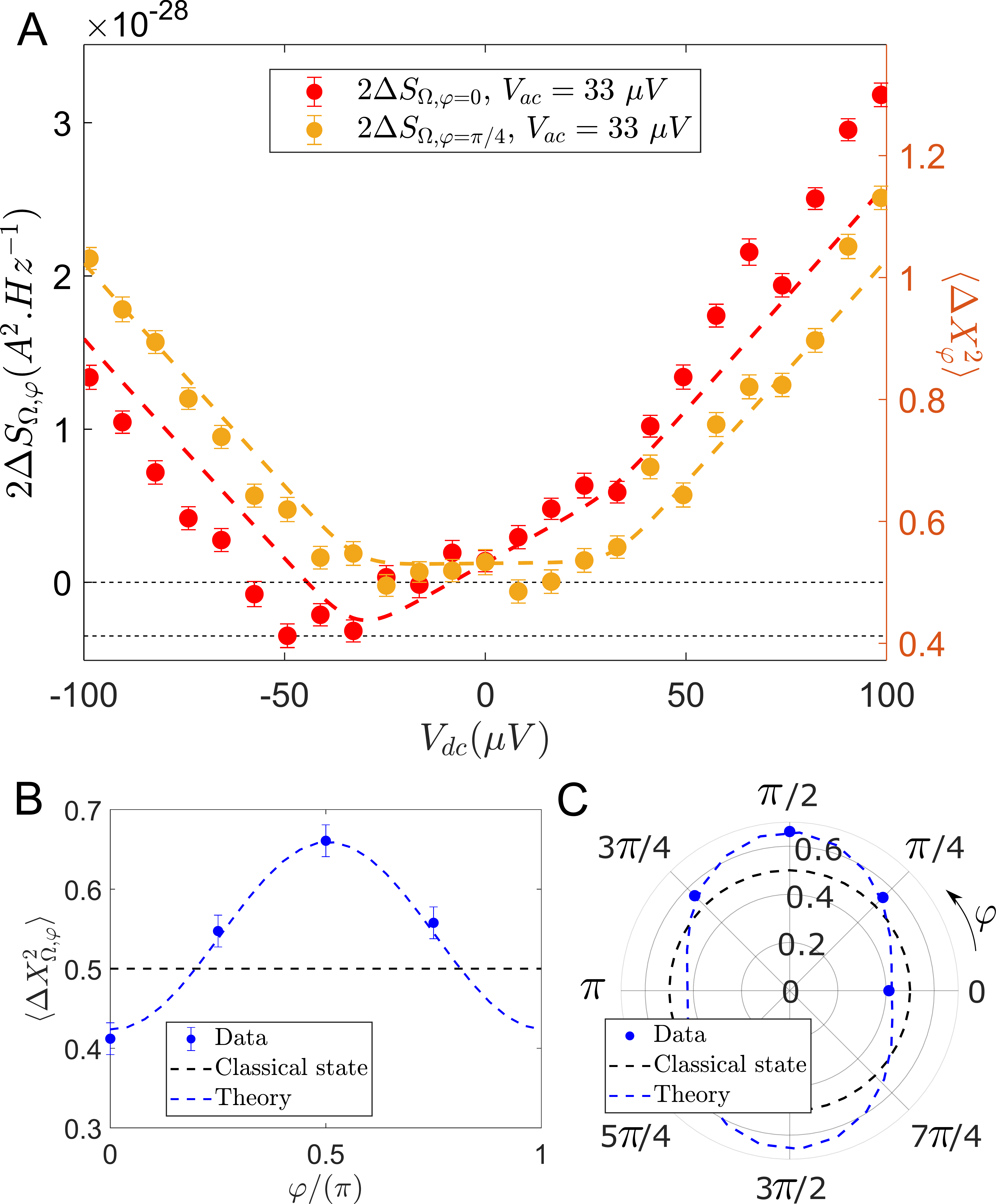}
\caption{\label{fig3}
{\bf A.}  Measurements of $\Delta S_{\Omega,\varphi}$ for $\varphi=0$ (red points) and $\varphi=\pi/4$ (yellow points) as a function of the dc bias voltage for $V_{\text{ac}}=33$ $\mu$V. The red and yellow dashed lines represent the theoretical predictions for $V_{\text{ac}}=33$ $\mu$V. Best agreement is obtained using $T_{\text{el}}=30$ mK slightly lower than $T_{\text{el}}=40$ mK which gave the best agreement for low frequency noise measurements. {\bf B.} Measurements of $\langle \Delta X_{\Omega,\varphi}^{2} \rangle $ as a function of $\varphi$ for $V_{\text{dc}}=-33$ $\mu$V. The black dashed line represents the classical fluctuations and the blue dashed line the theory. {\bf C.} Same data in polar coordinates. }
\end{figure}

We now turn to the noise measurements performed when the pump is on, with $V_{\text{ac}}=33$ $\mu$V. The output signal is then directly proportional to $ S_{\Omega,\varphi}(V_{\text{dc}}, V_{\text{ac}})- S_{\Omega,\varphi}(V_{\text{dc}}=0, V_{\text{ac}})$. In order to reconstruct the excess output noise, $\Delta S_{\Omega,\varphi}(V_{\text{dc}}, V_{\text{ac}})$ it is necessary to independently measure the excess high frequency noise generated by the pump: $\Delta S_{\Omega,\varphi}(V_{\text{dc}}=0, V_{\text{ac}})$. This is performed by measuring the excess noise when the pump amplitude is modulated at $f_m=234$ Hz at zero dc voltage. By summing these two contributions, one obtains $\Delta S_{\Omega,\varphi}(V_{\text{dc}}, V_{\text{ac}})=S_{\Omega,\varphi}(V_{\text{dc}}, V_{\text{ac}})- S_{\Omega,\varphi}(V_{\text{dc}}=0, V_{\text{ac}}) + \Delta S_{\Omega,\varphi}(V_{\text{dc}}=0, V_{\text{ac}})$. We discuss first the low-frequency measurements, $\Delta S_0(V_{\text{dc}},V_{\text{ac}})$, represented by the red dots in Figure 2. For $V_{\text{dc}}=0$, the excess low-frequency noise is set by the partitioning of electron-hole pairs generated by the pump. For $V_{\text{dc}}> hf/e$, we observe $\Delta S_0(V_{\text{dc}},V_{\text{ac}}) = \Delta S_0(V_{\text{dc}},0)$. This is due to two particle interferences occurring between the electrons emitted by the dc source and the electron-hole pairs generated by the ac pump which fully suppress the partition noise of the pump. Our data agree well with the red dashed line, which is the theoretical prediction for a pump amplitude $V_{\text{ac}}=33$ $\mu$V and $T_{\text{el}}=40$ mK \cite{Note2}.
The high frequency noise measurements plotted in Figure 3 show a completely different behavior. $\Delta S_{\Omega,\varphi}(V_{\text{dc}},V_{\text{ac}})$ depends strongly on the phase $\varphi$ as shown by the strong differences between $\varphi=0$ (red points) and $\varphi=\pi/4$ (yellow points). $\Delta S_{\Omega,\varphi=\pi/4}(V_{\text{dc}},V_{\text{ac}})$ resembles the measurement in the absence of the pump with a slight shift upwards corresponding to the partition noise of the pump. In particular, it is symmetric for positive and negative biases $V_{\text{dc}}$. $ \Delta S_{\Omega,\varphi=0}(V_{\text{dc}},V_{\text{ac}})$ looks completely different. It is asymmetric as a function of $V_{\text{dc}}$. It is larger for $V_{\text{dc}}>0$ compared to $V_{\text{dc}}<0$ and even goes below zero for $V_{\text{dc}}\approx -V_{\text{ac}}=-33 \mu$V. It shows that for this combination of dc and ac voltages, squeezing of the EMP mode at frequency $f$ is obtained. The observed asymmetry can be easily explained. Due to the electron-hole symmetry of the pump (the sine excitation is symmetric with respect to positive and negative energies), one has $ \Delta S_{\Omega,\varphi=0}(V_{\text{dc}},V_{\text{ac}}) = \Delta S_{\Omega,\varphi=\pi/2}(-V_{\text{dc}},V_{\text{ac}})$. Both quadratures for a given sign of the dc bias can thus be accessed from the measurement of a single quadrature for positive and negative bias. For $V_{\text{dc}} \approx - V_{\text{ac}}$, $\Delta S_{\Omega,\varphi=0}(V_{\text{dc}},V_{\text{ac}})$ is negative; the other quadrature $\Delta S_{\Omega,\varphi=\pi/2}(V_{\text{dc}},V_{\text{ac}})$ then shows an excess noise compared to the equilibrium situation as expected from Heisenberg uncertainty principle.

Finally, we can convert our noise measurements into the fluctuations $\langle \Delta X_{\Omega,\varphi}^{2} \rangle $ using Eq.(\ref{X2}). We have plotted in Figure 4.A $\langle \Delta X_{\Omega,\varphi}^{2} \rangle $ as a function of $\varphi$ for $V_{\text{dc}}=-33$ $\mu$V. In order to emphasize the anisotropy of the noise, we have plotted the same data in polar coordinates in Figure 4.B. The classical isotropic fluctuations are represented by the black dashed line and the theoretical predictions by the blue dashed line which agrees very well with our data. As discussed before, a clear squeezing can be observed for $\varphi=0$ with an $18\%$ reduction compared to vacuum fluctuations.

To conclude, we have demonstrated squeezing of EMP modes at frequency $f=7.75$ GHz by using two-particle interferences in an electronic beam-splitter between a dc and an ac sinusoidal electronic sources. Squeezed EMP states could be used in EMP interferometers for quantum enhanced sensors or in EMP cavities used as quantum buses to transmit quantum states between distant mesoscopic samples. For practical applications, it will be necessary to increase the degree of squeezing which could be achieved following two different ways. Firstly, one can structure the ac drive, replacing its sinusoidal temporal dependence by Lorentzian shaped current pulses\cite{Mendes2015,Ferraro2018,Rebora2021}. Secondly, the squeezing demonstrated here is based on the non-linear evolution of the high frequency noise with the applied dc bias voltage. Much larger non-linearities can be obtained using mesoscopic conductors such as ac driven mesoscopic capacitors \cite{Feve2007} for much larger squeezing efficiency \cite{Mendes2015}. The versatility of quantum Hall conductors can also be exploited by coupling non-linear mesoscopic conductors to quantum Hall resonators for parametric amplification and squeezing. Many building blocks of edge magnetoplasmonics are already available, the present work demonstrates that they can be combined in order to generate quantum EMP states.

We thank D. Ferraro and G. Rebora for useful discussions. This work is supported by the ANR grant "Qusig4Qusense", ANR-21-CE47-0012, the project EMPIR 17FUN04 SEQUOIA, and the French RENATECH network.

\end{document}


\captionsetup[figure]{labelformat=empty}

\title{Observation of edge magnetoplasmon squeezing in a quantum Hall conductor: supplementary material}
\author{ H. Bartolomei$^{1}$, R. Bisognin$^{1}$, H. Kamata$^{1}$,  J.-M. Berroir$^{1 }$,  E. Bocquillon$^{1,2 }$, G. M\'{e}nard$^{1 }$, B.
Pla\c{c}ais$^{1 }$,   A. Cavanna$^{3}$, U.
Gennser$^{3}$, Y. Jin$^{3}$,   P. Degiovanni$^{4}$, C. Mora$^{5}$, and G. F\`{e}ve$^{1 \ast}$ \\
\normalsize{$^{1}$ Laboratoire de Physique de l\textquoteright Ecole normale sup\'erieure, ENS, Universit\'e
PSL, CNRS, Sorbonne Universit\'e, Universit\'e Paris Cit\'e, Paris, France}\\
\normalsize{$^{2}$ II. Physikalisches Institut, Universit\"{a}t zu K\"{o}ln, Z\"{u}lpicher Str. 77, 50937 K\"{o}ln}\\
\normalsize{$^{3}$ Centre de Nanosciences et de Nanotechnologies (C2N), CNRS, Universit\'{e} Paris-Saclay, 91120 Palaiseau, France. }\\
\normalsize{$^{4}$ Univ  Lyon,  Ens  de  Lyon,  Universit\'e  Claude  Bernard  Lyon  1, CNRS,  Laboratoire  de  Physique,  F-69342  Lyon,  France}\\
\normalsize{$^{5}$ Universit\'e  de  Paris,  Laboratoire  Mat\'eriaux  et  Ph\'enom\`enes  Quantiques,  CNRS,  F-75013  Paris,  France.}\\
}

\maketitle

Our experimental measurements of squeezing can be compared with theoretical predictions using scattering theory of photo-assisted electron transport\cite{Pedersen98}. As shown in Refs.[\onlinecite{Bocquillon2014}] and [\onlinecite{Marguerite2017}], the current correlations in the time domain at the output of the QPC can be obtained from the input current correlations and from the incoming single electron coherence\cite{Grenier2011}. Going to frequency space, one can then extract the non stationary part, at pulsation $2\Omega$, of the time dependent current noise spectrum at $\Omega$, thus leading to the anisotropic contribution $\Delta S_{\Omega,\text{an}}$, whereas the time averaged current noise spectrum at $\Omega$ leads to the isotropic contribution $\Delta S_{\Omega,\text{iso}}$. In the experimental configuration considered here, input 1 is d.c. biased, such that the incoming single electron coherence in input 1 can be expressed as a function of the excess electronic distribution function $\Delta f_{\mu}(\Omega) =f_\mu(\omega) -f_0(\omega) $, where $f_\mu(\omega)$ is the Fermi-Dirac distribution for electrons at chemical potential $\mu=-eV$ and $f_0(\omega)$ is the Fermi-Dirac distribution at the reference chemical potential $\mu=0$. Input 2 being a.c. biased, the excess electronic coherence at input 2 can be expressed as a function of the excess electronic Wigner distribution function $\Delta W(t,\Omega)$ introduced in Ref.[\onlinecite{Ferraro2013}]. As a result, $\Delta S_{\Omega, \varphi}= \Delta S_{\Omega,\text{iso}} + \Delta S_{\Omega,\text{an}}$ is given by:

\begin{eqnarray}
\frac{\Delta S_{\Omega,\text{iso}}}{RT e^2} & = &  \int \frac{d\Omega'}{2\pi} \big[ \overline{\Delta W(t,\Omega')}^{t} g_{\mu}(\Omega,\Omega') + \Delta f_{\mu}(\Omega') g_{0}(\Omega,\Omega') \big] \nonumber \\
\label{DSis}\\
\frac{\Delta S_{\Omega,\text{an}}}{RT e^2} & = &  \Re\big[\int \frac{d\Omega'}{2\pi}  \overline{e^{2i\Omega t} \Delta W(t,\Omega')}^{t}(1- 2f_{\mu}(\Omega')) e^{2i\varphi} \big] \label{DSan}
\end{eqnarray}
in which  $g_\mu(\Omega,\Omega')=(1-f_\mu(\Omega'-\Omega)-f_\mu(\Omega'+\Omega))$. In the case where source 2 is a periodic a.c. drive at frequency $f$, $V_{\text{ac}}(t)$, the excess Wigner distribution $\Delta W(t,\omega)$ can be computed from the photo-assisted transition amplitudes $p_n[V_{\text{ac}}]$ defined as:

\begin{eqnarray}
e^{-i\frac{e}{\hbar} \int_{-\infty}^t V_{\text{ac}}(t')dt'} & = & \sum_n p_n[V_{\text{ac}}] e^{-2i \pi n f t}\\
 \Delta W(t,\omega) &  = & \sum_{n_+,n_-} p_{n_+}[V_{\text{ac}}]p_{n_-}[V_{\text{ac}}]^* e^{2i \pi (n_+-n_-) f t}   \nonumber  \\
 & \times & \big[f_{0}(\omega - \pi(n_++n_-) f) -f_0(\omega) \big]  \label{W}\\
\overline{\Delta W(t,\omega)}^{t} &= & \sum_{n} |p_{n}[V_{\text{ac}}]|^2 \big[f_{0}(\omega - 2\pi n  f) -f_0(\omega) \big] \nonumber \\
\end{eqnarray}

In the case of a sinusoidal a.c. drive, $V_{\text{ac}}(t) = V_{\text{ac}} \cos{(2\pi f t)}$, one has $p_{n}[V_{\text{ac}}] = J_n(eV_{\text{ac}}/(hf))$, where $J_n$ denotes the Bessel function of order $n$. The red and yellow dashed lines on Fig. S1 (Fig. 3 of the main manuscript) result from numerical evaluations of Eqs.(\ref{DSis}), (\ref{DSan}) and (\ref{W}).

\begin{figure}[h!]
\includegraphics[width=1
\columnwidth,keepaspectratio]{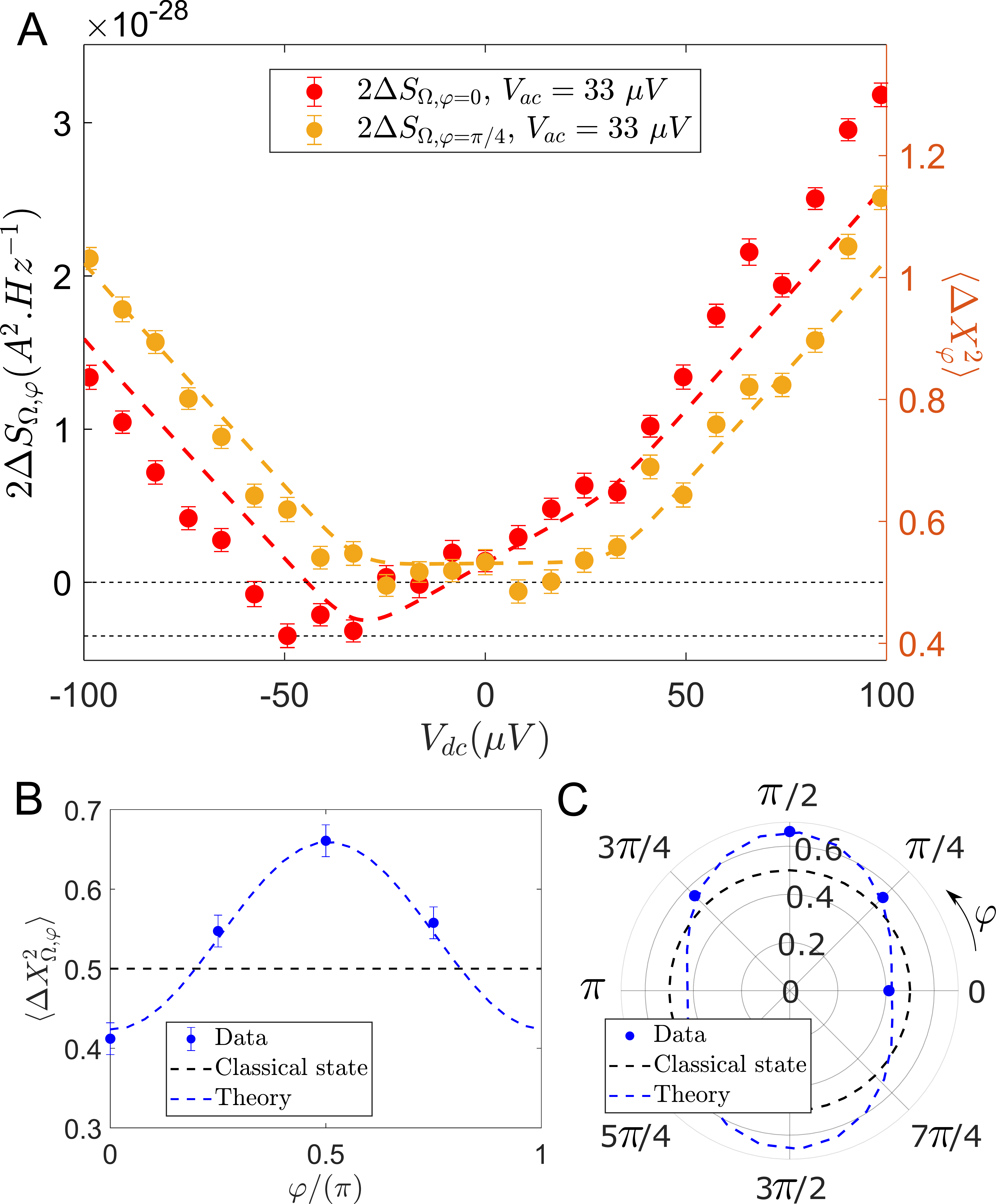}
\caption{\label{fig3}
{FIG. S1 \bf A.}  Measurements of $\Delta S_{\Omega,\varphi}$ for $\varphi=0$ (red points) and $\varphi=\pi/4$ (yellow points) as a function of the dc bias voltage for $V_{\text{ac}}=33$ $\mu$V. The red and yellow dashed lines represent the theoretical predictions for $V_{\text{ac}}=33$ $\mu$V obtained from numerical evaluations of Eqs.(\ref{DSis}), (\ref{DSan}) and (\ref{W}). Best agreement is obtained using $T_{\text{el}}=30$ mK slightly lower than $T_{\text{el}}=40$ mK which gave the best agreement for low frequency noise measurements. {\bf B.} Measurements of $\langle \Delta X_{\Omega,\varphi}^{2} \rangle $ as a function of $\varphi$ for $V_{\text{dc}}=-33$ $\mu$V. The black dashed line represents the classical fluctuations and the blue dashed line the theory. {\bf C.} Same data in polar coordinates. }
\end{figure}